\documentstyle[aps,12pt,preprint]{revtex}
\begin{document}
\title{The decays $\bar{B}\rightarrow \bar{K}D$ and $\bar{B}\rightarrow \bar{K}\bar{%
D}$ and final state interactions}
\author{Fayyazuddin}
\address{National Centre for Physics and Physics Department, Quaid-i-Azam University,%
\\
Islamabad, Pakistan.}
\maketitle

\begin{abstract}
The decays $\bar{B}\rightarrow \bar{K}D$ and $\bar{B}\rightarrow \bar{K}\bar{%
D}$ taking into account final state interactions are discussed. These decays
are described by four strong phases $\delta _0,\delta _1,\tilde{\delta}_0,%
\tilde{\delta}_1$(subscript 0 and 1 refer to $I=0$ and $I=1$ isospin final
states), one weak phase $\gamma $ and four real amplitudes. Isospin
constraints are taken into account. It is argued that strong interaction
dynamics gives $\delta _1\approx \tilde{\delta}_1$. The four real amplitudes
are estimated. Some observable consequences are discussed.

\newpage\ 
\end{abstract}

The weak decays $\bar{B}\rightarrow \bar{K}D$ and $\bar{B}\rightarrow \bar{K}%
\bar{D}$ taking into account final state interactions have been studied by
several authors [1,2,3,4]. In this paper we elaborate some of the points
discussed in reference [4]

These decays are described by four real amplitudes, four strong phases and
one weak phase $\gamma $. Since the effective weak Hamiltonian for these
decays has $\bigtriangleup I=1/2$, the isospin analysis give [4].

\begin{mathletters}
\begin{eqnarray}
A(\bar{B} &\rightarrow &K^{-}D^0)=2f_1e^{i\delta _1} \\
A(\bar{B}^0 &\rightarrow &K^{-}D^{+})=[f_1e^{i\delta _1}+f_0\text{ }%
e^{i\delta _0}] \\
A(\bar{B}^0 &\rightarrow &\bar{K}^0D^0)=[f_1e^{i\delta _1}-f_0\text{ }%
e^{i\delta _0}]
\end{eqnarray}

where $\delta _0$ and $\delta _1$are the phase shifts for $I=0$ and $I=1$
isospin states. On other hand for the decays $\bar{B}\rightarrow \bar{K}\bar{%
D}$, we have

\end{mathletters}
\begin{mathletters}
\begin{eqnarray}
A(\bar{B}^0 &\rightarrow &\bar{K}^0\bar{D}^0)=2\tilde{f}_1\text{ }e^{i\gamma
}\text{ }e^{i\tilde{\delta}_1} \\
A(B^{-} &\rightarrow &K^{-}\bar{D}^0)=e^{i\gamma }\text{ }[\tilde{f}_1\text{ 
}e^{i\tilde{\delta}_1}+\tilde{f}_0\text{ }e^{i\tilde{\delta}_0}] \\
A(B^{-} &\rightarrow &\bar{K}^0D^{-})=e^{i\gamma }\text{ }[-\tilde{f}_1\text{
}e^{i\tilde{\delta}_1}+\tilde{f}_0\text{ }e^{i\tilde{\delta}_0}]
\end{eqnarray}

>From Eqs. (1) and (2), we obtain

\end{mathletters}
\begin{equation}
R\equiv \frac{\Gamma (\bar{B}^0\rightarrow K^{-}D^{+})-\Gamma (\bar{B}%
^0\rightarrow \bar{K}^0D^0)}{\Gamma (\bar{B}^0\rightarrow K^{-}D^{+})+\Gamma
(\bar{B}^0\rightarrow \bar{K}^0D^0)}=\frac{2f_1f_0\cos (\delta _1-\delta _0)%
}{f_1^2+f_0^2}
\end{equation}

\begin{equation}
\tilde{R}\equiv \frac{\Gamma (B^{-}\rightarrow K^{-}\bar{D}^0)-\Gamma
(B^{-}\rightarrow \bar{K}^0D^{-})}{\Gamma (B^{-}\rightarrow \bar{K}%
^0D^{-})+\Gamma (\bar{B}\rightarrow \bar{K}^0D^{-})}=\frac{2\tilde{f}_1%
\tilde{f}_0\cos (\tilde{\delta}_1-\tilde{\delta}_0)}{\tilde{f}_1^2+\tilde{f}%
_0^2}
\end{equation}

Further if we consider the decay of $B^{\mp }$ to CP-eigenstates $%
D_{1,2}=(D^0\mp \bar{D}^0)/\sqrt{2}$ (in our convention $D_1$ and $D_2$ have 
$CP=+1$ and $-1$ respectively), we obtain

\[
R_{1,2}\equiv \Gamma (B^{-}\rightarrow K^{-}D_{1,2})+\Gamma
(B^{+}\rightarrow K^{+}D_{1,2})/\Gamma (B^{-}\rightarrow K^{-}D^0) 
\]

\begin{equation}
=[1+\frac 14(r_1^2+r_0^2+2r_1r_2\cos (\tilde{\delta}_1-\tilde{\delta}_0)\mp
r_1\cos \gamma \cos (\tilde{\delta}_1-\delta _1)\mp r_0\cos \gamma \cos (%
\bar{\delta}_1-\delta _1)]
\end{equation}

\begin{eqnarray}
{\cal A}_{1,2} &\equiv &\frac{\Gamma (B^{-}\rightarrow K^{-}D_{1,2})-\Gamma
(B^{+}\rightarrow K^{+}D_{1,2})}{\Gamma (B^{-}\rightarrow K^{-}D^0)} 
\nonumber \\
&=&\pm [r_1\sin \gamma (\tilde{\delta}_1-\delta _1)+r_0\sin \gamma \sin (%
\tilde{\delta}_0-\delta _1)]
\end{eqnarray}

where

\begin{equation}
r_1=\tilde{f}_1/f_1,r_0=\tilde{f}_0/f_1
\end{equation}

It may be noted that we get the result of reference [4] if we put $\tilde{f}%
_0=\tilde{f}_1$ and $\tilde{\delta}_0=\tilde{\delta}_1.$

So far our analysis is general. To proceed further we note that these decays
are determined by the tree amplitude $T$, the color suppressed amplitudes $C(%
\tilde{C})$ and annihilation amplitude $\tilde{A}.$ In terms of these
amplitudes

\begin{mathletters}
\begin{eqnarray}
f_1 &=&\frac{G_F}{\sqrt{2}}|V_{cb}V_{us}^{*}|\frac 12(T+C) \\
f_0 &=&\frac{G_F}{\sqrt{2}}|V_{cb}V_{us}^{*}|\frac 12(T-C)
\end{eqnarray}

and

\end{mathletters}
\begin{mathletters}
\begin{eqnarray}
\tilde{f}_1 &=&\frac{G_F}{\sqrt{2}}|V_{ub}V_{cs}^{*}|\frac 12\tilde{C} \\
\tilde{f}_0 &=&\frac{G_F}{\sqrt{2}}|V_{ub}V_{cs}^{*}|\frac 12(\tilde{C}+2%
\tilde{A})
\end{eqnarray}

In the Wolfenstein representation of CKM matrix [5]

\end{mathletters}
\begin{equation}
\left| \frac{V_{ub}V_{cs}}{V_{cb}V_{us}^{*}}\right| =\sqrt{\rho ^2+\eta ^2}
\end{equation}

Thus we get

\begin{equation}
r_1=\sqrt{\rho ^2+}\eta ^2\left( \frac{\tilde{C}}{T+C}\right)
\end{equation}

\begin{equation}
r_0=\sqrt{\rho ^2+}\eta ^2\left( \frac{\tilde{C}+2\tilde{A}}{T+C}\right)
\end{equation}

\begin{equation}
R=\frac{T^2-C^2}{T^2+C^2}\cos (\delta _1-\delta _0)\simeq (1-2\frac{C^2}{T^2}%
)\cos (\delta _1-\delta _0)
\end{equation}

\begin{equation}
\tilde{R}=\frac{2\tilde{C}(\tilde{C}+2\tilde{A})}{\tilde{C}^2+(\tilde{C}+2%
\tilde{A})^2}\cos (\tilde{\delta}_1-\tilde{\delta}_0)\approx (1-2\tilde{A}^2/%
\tilde{C}^2)\cos (\tilde{\delta}_1-\tilde{\delta}_0)
\end{equation}

where we have retained only the terms upto $C^2/T^2$ and $\tilde{A}^2/\tilde{%
C}^2,$ since $C^2/T^2$ and $\tilde{A}^2/\tilde{C}^2$ are small (see below).

The following remarks about the strong phases are in order. Consider the
S-wave scattering

\begin{equation}
\bar{K}+D\rightarrow \bar{K}+D
\end{equation}

\begin{equation}
\bar{K}+\bar{D}\rightarrow \bar{K}+\bar{D}
\end{equation}

Since $\bar{K}\sim s\bar{q}$ and $D\sim c\bar{q}$, $q=u$ or $d,$no $s$ and $%
u $ channels poles are allowed, whereas since $\bar{D}\sim q\bar{c}$, the
poles with the quantum number of $D_{so}^{-}$ are possible in these
channels. But these states carry $I=0$, hence these states will contribute
to $I=0$ scattering amplitude i.e. to $\bar{\delta}_0.$ The t channel is
common to the processes $(15)$ and $(16)$ and the lowest lying poles which
can contribute are $\rho $ and $\sigma $. The $\rho $ and $\sigma -$ pole,
contribute to $I=1$ and $I=0$ channel, respectively. Thus it is reasonable
to assume that $(\delta _1=\tilde{\delta}_1)$; since $s$ and $u$-channels
poles do not contribute to $I=1$ scattering amplitudes. Thus with $\delta _1=%
\tilde{\delta}_1$, we obtain from Eqs. $(5)$ and $(6)$

\begin{equation}
R_2-R_1=2\cos \gamma \left( r_1+r_0\cos (\tilde{\delta}_0-\tilde{\delta}%
_1)\right)
\end{equation}

\begin{equation}
{\cal A}_{1,2}=\pm \text{ }r_0\sin \gamma \sin (\tilde{\delta}_0-\tilde{%
\delta}_1)
\end{equation}

First we note that if $2(\tilde{A}/_{\tilde{C}})<<1$, then Eq. $(14)$ gives $%
\cos $ $(\tilde{\delta}_0-\tilde{\delta}_1)$ in term of $\tilde{R}$, and
then from Eq. $(18)$ one can extract $r_0\sin \gamma $. If $r_1\simeq r_0$
which is the case if $2(\tilde{A}/_{\tilde{C}})<<1$, then, Eqs. $(14),(17)$
and $(18)$ give us information about $r_0$ and the weak phase $\gamma $.

Before we give some estimates for the amplitudes $T,C(\tilde{C})$ and $%
\tilde{A}$, we discuss $SU(3)$ relations for the various amplitudes for the
decays of $\bar{B}$ described by the effective Lagrangians:

\begin{equation}
L_{eff}=\frac{G_F}{\sqrt{2}}V_{cb}V_{us}^{*}[\bar{s}\gamma _\mu (1+\gamma
_5)u][\bar{c}\gamma _\mu (1+\gamma _5)b]
\end{equation}

\begin{equation}
L_{eff}=\frac{G_F}{\sqrt{2}}V_{ub}V_{cs}^{*}[\bar{s}\gamma _\mu (1+\gamma
_5)c][\bar{u}\gamma _\mu (1+\gamma _5)b]
\end{equation}

$SU(3)$ analysis of these decays gives the following relations between
various amplitudes 
\begin{mathletters}
\begin{eqnarray}
A(B^{-} &\rightarrow &K^{-}D^0)=A(\bar{B}^0\rightarrow K^{-}D^{+})+A(\bar{B}%
^0\rightarrow \bar{K}^0D^0) \\
A(\bar{B}_s^0 &\rightarrow &\pi ^{-}D^{+})=\sqrt{2}A(\bar{B}_s^0\rightarrow
\pi ^0D^0) \\
\sqrt{6}A(\bar{B}_s^0 &\rightarrow &\eta D^{+})=A(\bar{B}_s^0\rightarrow \pi
^{-}D^{+})-2A(\bar{B}_s^0\rightarrow \bar{K}^0D^0)
\end{eqnarray}
and

\end{mathletters}
\begin{mathletters}
\begin{eqnarray}
A\left( B^{-}\rightarrow K^{-}\bar{D}^0\right) &=&A(B^{-}\rightarrow \bar{K}%
^0D^{-})+A(\bar{B}^0\rightarrow \bar{K}^0\bar{D}^0) \\
A\left( \bar{B}^0\rightarrow \pi ^{+}D_s^{-}\right) &=&\sqrt{2}A\left(
B^{-}\rightarrow \pi ^0D_s^{-}\right) \\
A\left( \bar{B}_s^0\rightarrow \pi ^{+}D^{-}\right) &=&\sqrt{2}A\left( \bar{B%
}_s^0\rightarrow \pi ^0D^0\right) \\
\sqrt{6}A\left( B^{-}\rightarrow \eta D_s^{-}\right) &=&A(\bar{B}%
^0\rightarrow \pi ^{+}D_s^{-})-2A(B^{-}\rightarrow \bar{K}^0D^{-}) \\
\sqrt{6}A\left( \bar{B}_s^0\rightarrow \eta \bar{D}^0\right) &=&A(\bar{B}%
_s^0\rightarrow \pi ^{+}D^{-})-2A(\bar{B}^0\rightarrow \bar{K}^0\bar{D}^0)
\end{eqnarray}

It may be noted that relations $(21,a,b)$ and $(22a,b,c)$ also follows from
isospin analysis only.

We now calculate the amplitudes, $T,C(\tilde{C})$ and $\tilde{A}$ from the
effective Lagrangians $(19)$ and $(20)$. In the factorization anstz, they
are given by

\end{mathletters}
\begin{eqnarray}
T &=&a_1f_KF_0^{B-D}(m_K^2)(m_B^2-m_D^2) \\
C &=&a_2f_DF_0^{B-K}(m_D^2)(m_B^2-m_K^2)=\tilde{C} \\
\tilde{A} &=&a_1f_BF_0^{D-K}(m_B^2)(m_D^2-m_K^2)
\end{eqnarray}

where the form factor F$_0$ is defined as $[t=(p-p^{\prime })^2]$ $(q=c$ or $%
u)$%
\begin{eqnarray}
&&\left\langle P(p^{\prime })\left| i\bar{q}\gamma _\mu (1+\gamma
_5)b\right| B(p)\right\rangle  \nonumber \\
&\sim &\left[ F_{+}(t)(p+p^{\prime })_\mu +F_{-}(t)(p-p^{\prime })_\mu
\right]  \nonumber \\
&=&\left[ (p+p^{\prime })_\mu -\frac{m_B^2-m_P^2}t(p+p^{\prime })_\mu
\right] F_1(t)  \nonumber \\
&&+\left[ \frac{m_B^2-m_P^2}t(p-p^{\prime })_\mu \right] F_0(t)
\end{eqnarray}

One can get some information for the form factor $F_0^{B-D}(t)$ from the
heavy quark effective theory $[6]$, but for the form factors involving one
light meson we use a model which is based on dispersion relations, using
once-subtracted dispersion relation for $[F_{+}(t)+F_{-}(t)]$ and
unsubtracted dispersion relatiuon for $[F_t(t)+F_{-}(t)]$ as given in
reference $[7]$. Retaining only the contribution from the low lying states $%
B^{*}(1^{-})$ and $B_0(O^{+})$ in the dispersion relations, one gets

\begin{mathletters}
\begin{eqnarray}
F_1(t) &=&\frac 12\left[ \frac{f_B}{f_P}-\frac{f_{B^{*}}g_{B^{*}BP}}{%
m_{B^{*}}}-\frac{f_{B_0}g_{B_0BP}}{m_{B_0}}+\frac{2f_{B^{*}}g_{B^{*}BP}}{%
m_{B^{*}}^2-t}\right] \\
F_0(t) &=&\frac 12\left\{ \left[ \frac{f_B}{f_P}-\frac{f_{B^{*}}g_{B^{*}BP}}{%
m_{B^{*}}}-2f_{B_0}g_{B_0BP}\frac{m_{B_0}}{m_B^2}+\frac{f_{B_0}g_{B_0BP}}{%
m_{B_0}}\right] \right.  \nonumber \\
&&+\frac t{m_B^2}\left[ \frac{f_B}{f_P}-\frac{f_{B^{*}}g_{B^{*}BP}}{m_{B^{*}}%
}\right]  \nonumber \\
&&+2\left. \left[ \frac{m_{B_0}}{m_B^2}f_{B_0}g_{B_0BP}\frac{m_{B_0}^2-m_B^2%
}{m_{B_0}^2-t}\right] \right\}
\end{eqnarray}

If we demand that there should not be a term depending linearly on $t$ in $%
F_0(t)$, we get the sum rule $[8]$

\end{mathletters}
\begin{equation}
\frac{f_B}{f_P}=\frac{f_{B^{*}}g_{B^{*}BP}}{m_{B^{*}}}+\frac{f_{B_0}g_{B_0BP}%
}{m_{B_0}}
\end{equation}

On using Eq. $(28)$, we obtain from Eqs. $(27)$, simple expressions for $%
F_1(t)$ and $F_0(t):$

\begin{equation}
F_1(t)=\frac{f_{B^{*}}m_{B^{*}}g_{B^{*}BP}}{m_{B^{*}}^2-t}
\end{equation}

\begin{equation}
F_0(t)=\left[ \frac{f_B}{f_P}-\frac{m_{B_0}^2}{m_{B_0}^2}(1-\frac{%
m_{B_0}^2-m_B^2}{m_{B_0}^2-t})(\frac{f_B}{f_P}-\frac{f_{B^{*}}g_{B^{*}BP}}{%
m_{B^{*}}})\right]
\end{equation}

Paramaterising

\begin{equation}
g_{B^{*}BP}=\lambda _B\frac{m_{B^{*}}}{f_P}
\end{equation}

and using the relation $[9]$

\begin{equation}
f_{B^{*}}=f_B
\end{equation}

we get

\begin{equation}
F_1(t)=\lambda _B\frac{f_B}{f_P}\frac{m_{B^{*}}^2}{m_{B^{*}}^2-t}
\end{equation}

\begin{equation}
F_0(t)=\frac{f_B}{f_P}\left\{ \lambda _B-\frac{m_{B_0}^2-m_B^2}{m_B^2}%
(1-\lambda _B)\left[ 1-\frac{m_{B_0}^2}{m_{B_0}^2-t}\right] \right\}
\end{equation}

\begin{equation}
F_1(0)=F_0(0)=\lambda _B\frac{f_B}{f_P}
\end{equation}

These form factors except for $\lambda _B$ depend upon the ratio $\frac{f_B}{%
f_P}$ and masses $m_B,m_{B^{*}}$ and $m_{B_0}$ which can be extracted from
the experimental data. We assume that $\lambda _B$ and $\lambda _D$ scale as:

\begin{equation}
\lambda _B=\frac \Lambda {m_B},\lambda _D=\Lambda /m_D
\end{equation}

where $\Lambda $ is a scale characteristic of bound state which we take 1
GeV. Using this assumption, we get

\begin{eqnarray}
\Gamma \left( D^{*}\rightarrow D^0\pi ^{+}\right) &=&\frac{g_{D^{*}D\pi }^2}{%
6\pi }\frac{p_\pi ^3}{m_{D^{*}}^2}  \nonumber \\
&=&\frac 1{6\pi }\left( \frac \Lambda {m_D}\frac{m_{D^{*}}}{f_\pi }\right) ^2%
\frac{p_\pi ^3}{m_{D^{*}}^2}  \nonumber \\
&=&\frac 1{6\pi }\left( \frac \Lambda {m_D}\right) ^2\frac{p_\pi ^3}{f_\pi ^2%
}\approx 52KeV
\end{eqnarray}

Thus we obtain

\begin{equation}
\Gamma (D^{**}\rightarrow D\pi )=\Gamma (D^{*+}\rightarrow D^0\pi
^{+})+\Gamma (D^{*+}\rightarrow D^{+}\pi ^0)\simeq 78\text{ KeV}
\end{equation}

to be compared with the experimental upper limit $[10]$ $\Gamma <113$ KeV.
Thus our assumption that $\lambda _D=\Lambda /m_D$ can be tested
experimentally. Finally using Eqs. $(36)$, we get 
\begin{eqnarray}
F_0^{D-K}(m_B^2) &=&\frac{f_B}{f_K}\left[ \frac \Lambda {m_D}-\frac{%
m_{B_0}^2-m_B^2}{m_B^2}(1-\frac \Lambda {m_D})(1-\frac{m_{B_0}^2}{%
m_{B_0}^2-m_D^2})\right] \\
F_0^{D-K}(m_B^2) &=&\frac{f_{D_s}}{f_K}\left[ \frac \Lambda {m_{D_s}}-\frac{%
m_{s_0}^2-m_D^2}{m_D^2}(1-\frac \Lambda {m_{D_s}})(1-\frac{m_{D_{s_0}}^2}{%
m_{D_{s_0}}^2-m_B^2})\right]
\end{eqnarray}

Using following values for the masses (in $GeV$): $[11]$ $%
m_B=5.279,m_{B_0}=5.60,m_{D_s}=1.968,m_{D_{s_0}}=2.357,m_D=1.869$ and $%
f_D=200MeV,f_{D_s}=240MeV,f_B=180MeV$ and $f_K=158MeV[6],$ we obtain 
\begin{eqnarray}
F_0^{B-K}(m_B^2) &\simeq &0.202\text{ }\frac{f_B}{f_K}\simeq 0.23 \\
F_0^{D-K}(m_B^2) &\simeq &0.145\text{ }\frac{f_{D_S}}{f_K}\simeq 0.22
\end{eqnarray}

Hence we get 
\begin{eqnarray}
\tilde{A}/\tilde{C} &=&\frac{a_1}{a_2}\text{ }\frac{f_B}{f_K}\text{ }\frac{%
F_0^{D-K}(m_B^2)}{F_0^{B-K}(m_D^2)}\left( \text{ }\frac{m_D^2-m_K^2}{%
m_B^2-m_K^2}\right) \simeq 0.120 \\
C/T &=&\frac{a_2}{a_1}\text{ }\frac{f_DF_0^{B-K}(m_D^2)(m_B^2-m_K^2)}{%
f_KF_0^{D-K}(m_K^2)(m_B^2-m_D^2}\simeq 0.126
\end{eqnarray}
where we have used for the color suppression factor [12] 
\begin{equation}
\frac{a_2}{a_1}=0.26
\end{equation}

and [6]

\begin{equation}
F_0^{B-D}(m_K^2)=0.587
\end{equation}

Now using Eqs. (43), and (44), and $\sqrt{\rho ^2+\eta ^2}\simeq 0.36[13]$
we get from Eqs. (11), (12), (13), (14), (17) and (18) 
\begin{eqnarray}
r_1 &\approx &0.040,r_0\approx 0.050,r_0/r_1=1.25 \\
R &\approx &0.968\cos (\delta _1-\delta _0) \\
\tilde{R} &\approx &0.971\cos (\tilde{\delta}_1-\tilde{\delta}_0) \\
R_2-R_1 &\approx &0.8\cos \gamma [1.25+\cos (\tilde{\delta}_1-\tilde{\delta}%
_0)] \\
{\cal A}_{1,2} &\approx &\pm 0.05\sin \gamma \sin (\tilde{\delta}_1-\tilde{%
\delta}_0)
\end{eqnarray}

If the neglect the term $2(\tilde{A}/\tilde{C})^2$, then we have 
\begin{eqnarray}
\cos (\tilde{\delta}_1-\tilde{\delta}_0) &\approx &\tilde{R} \\
{\cal A}_{1,2} &\approx &\pm (0.05)\sqrt{1-\tilde{R}^2}\sin \gamma \\
R_2-R_1 &\approx &0.08\cos \gamma [1.25+\tilde{R}]
\end{eqnarray}

To conclude if $2(\tilde{A}/\tilde{C})^2$ is negligible then the branching
ratio $\tilde{R}$, the asymmetry ${\cal A}_{1,2}$ and $R_2-R_1$ can give
information about the weak phase $\gamma $. Conversely if weak phase is
known from some other processes, $Eqs.(53),(54)$ and $\tilde{R}$ give us
information about r$_0$ and $r_0/r_1.$


\begin{references}
\bibitem{}  M. Gronau, Phys. Rev. D {\bf 58}, 037301 (1998).

\bibitem{bibitem}  M.Gronau and D Wyler, Phys. Lett. B {\bf 265}, 172,
(1991).

\bibitem{}  Z-Z. Xing, hep.ph/9804434, Phys. Rev. D.

\bibitem{bibitem}  M. Gronau and J.L. Rosner, hep-ph/9807447.

\bibitem{bibitem}  L. Wolfenstein, Phys. Rev. Lett. {\bf 51}, 1945 (1983);
N. Cabbibo, Phys. Rev. Lett. {\bf 10}, 531 (1963); M. Kobayski and K.
Maskawa, Prog. Theor. Phys. {\bf 49}. 652 (1973).

\bibitem{bibitem}  See for instance, H. Neubart and B. Stech, ''Non-leptonic
weak decays of B-mesons'' hep-ph 9705292; To appear in the Second Edition of
Heavy Flavors, edited by A.J. Buras and M. Linder (World Scientific,
Singapore).

\bibitem{bibitem}  Fayyazuddin and Riazuddin, Phys. Rev. D {\bf 49}, 3385
(1994). In the present paper we have used slightly different definitions for
g$_{B_0B\pi }$ and $f_{B^{*}}i.e.\frac{m_{B_0}^2-m_B^2}{2m_B\text{ }}%
g_{B_0B\pi }\rightarrow \frac{m_{B_0}^2-m_B^2}{m_{B_0}}g_{B_0B\pi
},f_{B^{*}}\rightarrow m_{B^{*}}f_{B^{*}}$

\bibitem{bibitem}  C.A. Domingues and N. Paver, Z. Phys. C {\bf 41}, 270
(1988).

\bibitem{bibitem}  N. Isgur and M.B. Wise, Phys. Lett. B {\bf 232, 113}
(1989). and {\bf 237}, 527 (1990).

\bibitem{bibitem}  Particle Data Group, Review of Particle Properties, Z.
Phys. C {\bf 3}, 620 (1998).

\bibitem{bibitem}  For the masses m$_{B_0}$ and m$_{D_{s_0}}$, see for
example Fayyazuddin and Riazuddin, J. Phys. G: Nucl. Part. Phys. {\bf 24, 23}
(1998).

\bibitem{bibitem}  T.E. Browder, K. Houscheid and D. Pedrini, Ann. Rev.
Nucl. Part. Sci. {\bf 46, 395} (1997).

\bibitem{bibitem}  A. Ali and D. London hep-ph/9707251
\end{references}
\end{document}